\documentclass[letterpaper]{article} 
\usepackage{aaai25}
\usepackage{times}  
\usepackage{helvet}  
\usepackage{courier}  
\usepackage[hyphens]{url}  
\usepackage{graphicx} 
\usepackage{natbib}  
\usepackage{caption} 
\usepackage{algorithm}
\usepackage{algorithmic}
\usepackage{subcaption}  
\usepackage[capitalize,noabbrev]{cleveref}
\usepackage{newfloat}
\usepackage{listings}
\DeclareCaptionStyle{ruled}{labelfont=normalfont,labelsep=colon,strut=off} 
\floatstyle{ruled}
\newfloat{listing}{tb}{lst}{}
\floatname{listing}{Listing}
\usepackage{todonotes}

\usepackage{xspace}
\newcommand{\xai}{\textsc{XAI}\xspace}
\newcommand{\rex}{\textsc{ReX}\xspace}
\newcommand{\threedrex}{\textsc{3D-ReX}\xspace}
\newcommand{\ie}{\emph{i.e}\xspace}
\newcommand{\eg}{\emph{e.g.}\xspace}

\title{3D ReX: Causal Explanations in 3D Neuroimaging Classification}

\author{
    Melane Navaratnarajah\textsuperscript{\rm 1},
    Sophie A. Martin\textsuperscript{\rm 2},
    David A. Kelly\textsuperscript{\rm 1},
    Nathan Blake\textsuperscript{\rm 1, \rm 2},
    Hana Chockler\textsuperscript{\rm 1}
}
\affiliations{
    \textsuperscript{\rm 1} King's College London, UK\\
    \textsuperscript{\rm 2} University College London, UK\\
    melane.navaratnarajah@kcl.ac.uk,
    s.martin.20@ucl.ac.uk,
    david.a.kelly@kcl.ac.uk,
    nathan.blake@kcl.ac.uk, 
    hana.chockler@kcl.ac.uk
    
}

\begin{document}

\maketitle

\begin{abstract}
Explainability remains a significant problem for AI models in medical imaging, making it challenging for clinicians to trust AI-driven predictions. We introduce \threedrex, the first causality-based post-hoc explainability tool for 3D models. \threedrex uses the theory of actual causality to generate responsibility maps which highlight the regions most crucial to the model's decision. We test \threedrex on a stroke detection model, providing insight into the spatial distribution of features relevant to stroke. 
\end{abstract}

\section{Introduction}

Machine learning, particularly deep learning algorithms, has transformed medical diagnosis by automating time-consuming and often subjective tasks, leading to more efficient and accurate results \cite{esteva2019guide}. These algorithms can extract features from complex data that may not be distinguishable to the human eye, reshaping healthcare practices. In particular, 3D data provides in-depth spatial information of internal structures which is often critical to understanding complex regions of the body. Spatial insights potentially aid with earlier diagnosis, facilitate intervention planning, modeling biological processes, and enable clinicians to visualize precise anatomy with greater reliability. However, despite the increasing use of 3D models across healthcare, a significant drawback is the lack of interpretability~\cite{SurveyMedicalXAI, saeed2023explainable}. If clinicians cannot discern the reasoning behind a model's final predictions, this poses a significant challenge for integrating these tools into medical practice. There is uncertainty and potential risks, as decisions based solely on opaque model outputs might have severe consequences, especially in a clinical setting. 

Explainable AI (XAI) tools, including perturbation-based methods like \textsc{Lime} ~\cite{LIME}, \textsc{Shap}~\cite{shap}, and \textsc{Rise}~\cite{petsiuk2018rise}, as well as backpropagation-based techniques such as Grad-CAM~\cite{CAM}, are designed to bridge the interpretability gap in AI-powered tools. These methods aim to facilitate understanding, increase trust in AI systems, and support error identification, paving the way towards safer and more transparent applications in high-risk domains such as healthcare. To this end, much attention has been given to XAI in diagnostic neuroimaging. 

However, existing methods still require substantial refinement in their robustness and adaptability to diverse input types, such as 3D data, to ensure they are suitable for explaining local and global disease markers. In 3D imaging, where identifying localized regions (\eg lesions) and broader patterns is critical, a notable gap exists in tools capable of delivering holistic explanations. This disparity remains largely under-explored.

Attempts to provide post-hoc explainability for 3D medical imaging models have typically focused on back-propagation techniques. For instance, \citet{wood2022deep} used smooth guided backpropagation to provide slice-wise and voxel-wise saliency maps. However, these methods rely upon direct access to the internal model, which is not always available. Hence, we focus here on model-agnostic XAI tools, which require no such access.

Much of the literature pertaining to XAI for neuroimaging has focused on cancer and dementia, as these are more clinically immediate problems \cite{van2022explainable}. However, a common limitation of these studies is the difficulty of validating the explanations at an individual level, due to the lack of a ground truth. Stroke detection, however, provides something against which to evaluate explanations: the presence of localized lesions in the brain. Of the few studies that apply explainability to stroke detection~\cite{gurmessa2023comprehensive}, the most common XAI methods are \textsc{SHAP} and \textsc{Lime}. Neither tool has an innate understanding of 3D data, nor of the types of occlusions required to extract maximum information from the model.

To overcome this, we introduce \threedrex, an extension of the existing framework, \rex~\cite{chockler2024causal}. \threedrex is modified to accommodate 3D inputs (without access to a model's weights) and to facilitate meaningful occlusion values. We present the algorithm for computing 3D causal explanations in~\Cref{sec:method} 
and conduct a small-scale evaluation of the method on stroke data (\Cref{sec:experiments}). We conclude with a discussion of present limitations and future work (\Cref{sec:discuss}).

\section{Related Work}\label{sec:relwork}

\rex, and \threedrex, are rooted in the framework of actual causality introduced by~\citep{HP05}. A thorough overview of the topic can be found in~\citet{halpern}.
The \rex tool is described in detail in~\citet{chockler2024causal}, and a full presentation of the theoretical apparatus is in~\citet{chockler2024explaining}. Actual causality extends simple counterfactual reasoning by considering the effect of \emph{interventions} to the current setting. For images, this means changing the values of pixels to some masking value, or values, and observing the output (\ie the model classification). 
\rex uses \emph{causal responsibility}~\cite{CH04} to quantify the contribution of pixels towards the final classification. We can view this quantification directly, as in~\Cref{fig:overlay}, or extract an approximately minimal set of pixels (or voxels) which are, by themselves, sufficient to get the required classification.

What constitutes a ``meaningful'' occlusion is open to debate and is often dependent on the dataset, task and model in question. \citet{fong2017interpretable} suggest that using naturalistic or plausible effects leads to more meaningful occlusions and experimented with constant values, noise injection and blurring. However, \citet{uzunova2019interpretable} and \citet{lenis2020domain} find that these occlusions are not suitable for medical imaging and instead introduced occlusion values using a variational autoencoder and inpainting respectively.~\citet{blake2024explainable} observed that \rex, with an occlusion value of $0$ on false RGB MRI slices of brain tumors, produces explanations which coincide well with a human provided segmentation.

\section{Methodology}%
\label{sec:method}

In this section, we present the details of \rex's inner workings, highlighting the modifications needed to accommodate 3D input data for medical imaging tasks. The algorithm functions by iteratively building a responsibility map that keeps track of each voxel's causal responsibility. This is achieved by masking specific input regions and assessing the model output for each mask. Occlusions are sets of supervoxels (themselves groupings of contiguous voxels): Responsibility for the classification is distributed over the supervoxels which contributed to the correct classification. No responsibility is given to those supervoxels which make no contribution. Supervoxels with non-$0$ responsibility are further refined. This process is repeated multiple times with different random supervoxels. The final result is a causal responsibility map. A sufficient, approximately minimal, subset of the voxels can be extracted from the image using information contained in the map, which constitutes a (causal) explanation. We highlight this explanation in orange in the examples shown in~\Cref{fig:overlay}.

The process is outlined for one iteration in~\Cref{fig:general-algo} and begins by obtaining the model's original prediction on an unmodified MRI scan, which serves as a \emph{target}. All subsequent predictions on modified inputs are compared to this target to determine whether a modification ``passes'' or ``fails'' — that is, whether it maintains the original classification. The responsibility map is initialized as a zero matrix matching the dimensions of the input data. This map accumulates responsibility for each voxel in the input space based on the passing mutants, eventually forming peaks in areas responsible to the classification. 

Each iteration of \threedrex involves initializing a queue that tracks partitions of the input space. Each supervoxel is attributed responsibility after querying the model, based on whether it was a passing partition. Supervoxels with no responsibility are discarded. This process continues until the search budget is exhausted or the supervoxels become so refined that they no longer contribute to the classification or fall below a size threshold.

Initially, the input is split into quadrants. We randomly select two axes (from $x$, $y$, and $z$) to partition the input space, resulting in four segments that can still adequately capture (or break) spatial dependencies in the 3D input. While dividing the input into eight partitions would provide more exhaustive coverage, it would lead to up to $256$ possible combinations to assess. In a worst-case scenario where all initial mutants pass, \ie a truly global diagnosis, each mutant would pass, and therefore, further generate $256$ additional variants, resulting in exponential growth of evaluation candidates. By reducing the partitioning to four segments as shown in \Cref{fig:gen-masks}, \threedrex constructs a responsibility map, highlighting causally significant regions in the 3D MRI data while reducing overwhelming computational costs. 



\begin{algorithm}[t]
    \centering
    \begin{algorithmic}[1]
        \STATE \textbf{Input:} Model $M$, Dataset $D$, Max Depth $d_{max}$, Search Limit $l_{max}$
        \STATE \textbf{Output:} Responsibility Map $rm$
        \STATE $m \leftarrow M$ \textbf{// Load the model with a specific checkpoint}
        \STATE $d \leftarrow D$ \textbf{// Select a valid input datapoint}
        \STATE $rm \leftarrow zeros(shape(d))$ \textbf{// Initialize Responsibility Map}
        \STATE $target \leftarrow m(d)$ \textbf{// Get classification for original datapoint}
        \STATE $queue \leftarrow [d]$ \textbf{// Initialize queue with the full input space}
        \STATE $passing\_mutants \leftarrow []$
        \STATE $total\_work \leftarrow 0$
        \WHILE{\textbf{not} ($depth \geq d_{max}$ \textbf{or} $steps \geq l_{max}$)}
            \STATE $head \leftarrow queue.pop()$
            \STATE $masks \leftarrow generate\_masks(head)$ \textbf{// Split input region into subregions}
            \FORALL{$mask \in masks$}
                \STATE $mutant \leftarrow apply\_mask(d, mask)$
                \STATE $prediction \leftarrow m(mutant)$
                \IF{$prediction = target$}
                    \STATE $queue.append(mask)$ \textbf{// Store passing mutants for further splitting}
                    \STATE $update\_resp\_map(rm, mask)$
                \ENDIF
            \ENDFOR
            \STATE $depth = update\_depth(depth, passing)$
            \STATE $total\_work += len(passing)$
        \ENDWHILE
        \STATE \textbf{return} $rm$
    \end{algorithmic}
    \caption{General Algorithm for \threedrex - $generate\_resp\_map()$}
    \label{fig:general-algo}
\end{algorithm}

\begin{algorithm}[t]
    \centering
    \begin{algorithmic}[1]
        \STATE \textbf{Input:} The head of the queue, $h$, which is a partition of the input. 
        \STATE \textbf{Output:} It will output the input randomly partitioned into 4 segments as a list.

        \STATE $axes \leftarrow [ROW, COL, DEPTH]$ 
        \textbf{// Define enums for the available axes}
        \STATE $ranges[ROW] \leftarrow [row\_start, row\_end]$
        \STATE $ranges[COL] \leftarrow [col\_start, col\_end]$
        \STATE $ranges[DEPTH] \leftarrow [depth\_start, depth\_end]$
        
        \STATE $axis1, axis2 \leftarrow rand(axes, 2)$ 
        \textbf{// Randomly picks two axes to split} 
        
        \STATE $coord1 \leftarrow rand(ranges[axis1])$
        \STATE $coord2 \leftarrow rand(ranges[axis2]) $
        
        \STATE $first\_split\_boxes = split(input\_box, axis1, coord1)$
        \FORALL{$box \in first\_split\_boxes$}
            \STATE $new\_box \leftarrow split(box, axis2, coord2)$
            \STATE $second\_split\_boxes.append(new\_boxes)$
        \ENDFOR
        \STATE \textbf{return} $second\_split\_boxes$
    \end{algorithmic}
    \caption{The mask generation algorithm: generate\_masks(h)}
    \label{fig:gen-masks}
\end{algorithm}

\section{Experiments}\label{sec:experiments}
While the purpose of this paper is to demonstrate \threedrex, we summarize here the data, model architecture and training. We trained a model to classify stroke patients from healthy controls using data from the Anatomical Tracings of Lesions After Stroke (ATLAS) R2.0~\cite{liew_large_2022} and IXI datasets~\cite{imperial_college_london_ixi_nodate}. 3D isotropic T1-weighted magnetic resonance images (MRI) from a total of $1236$ ($655$ stroke patients, $581$ controls) participants were used. Since data from the ATLAS dataset was already registered to a common brain template, we registered the IXI scans to the same template provided to reduce site-specific spatial biases. All volumes were skull-stripped, scaled to an intensity range of $0$ to $1$ and resized to 
$96\times 96 \times 96$mm. $90\%$ was used for training and validation, and $10\%$ as a held-out test set. We used MONAI's~\cite{monai_monai_2020} ResNet18 implementation to train a classifier using a binary cross-entropy loss with a weighted Adam optimizer. The trained model achieved an accuracy of $97\%$ and AUROC value of $99\%$ on the test data ($n=124$, $57$ controls and $67$ stroke patients).



\paragraph*{Settings for \threedrex Explanations}
To explain the trained model, we utilized the outlined method, \threedrex. To find the optimal occlusion value for examining the model's decision-making process in this specific context, we compared different options, such as the mean intensity across stroke patients, patches from a healthy MRI scan and a 0 value. This value is applied to the data after it is processed for the model, so does not in general correspond to black. Rather, it is an out-of-distribution value unlikely to be associated with a specific physical part of the brain scan.

While both healthy MRI occlusion and 0-value occlusion successfully produced explanations, mean value occlusion failed to yield any interpretable results, highlighting its ineffectiveness in this context. The explanations derived from the healthy MRI and 0-value occlusion highlighted distinct regions of the brain. The 0-value occlusion produced explanations closely aligned/overlapping with lesion location, indicating that the model's focus was on areas corresponding to regions of interest typically associated with stroke pathology. In contrast, the healthy MRI occlusion led to explanations that pointed to a different area of the brain, not aligned with lesion location but rather with other regions that the model had focused on during its decision-making process. 

The divergence in explanation for different occlusion values highlights an important insight: while the 0-value occlusion produced results that were closer to the lesion location, using the healthy MRI scan revealed that the model's underlying reasoning might rely on a complex set of brain regions for the classification. Although interesting, further experimentation on a larger model and dataset is required to confirm and validate these results. Nonetheless, these findings draw importance to the selection of the occlusion values and how using different strategies provides valuable insight into the model's interpretation of the stroke-related features in MRI data. From a clinician's perspective, being able to learn a diverse set of features that do not solely focus on the obvious pathological regions provides an opportunity to discover nuanced patterns and derive new biological insights. 

\begin{figure*}[hbt!]
    \centering
    \includegraphics[width=0.8\textwidth]{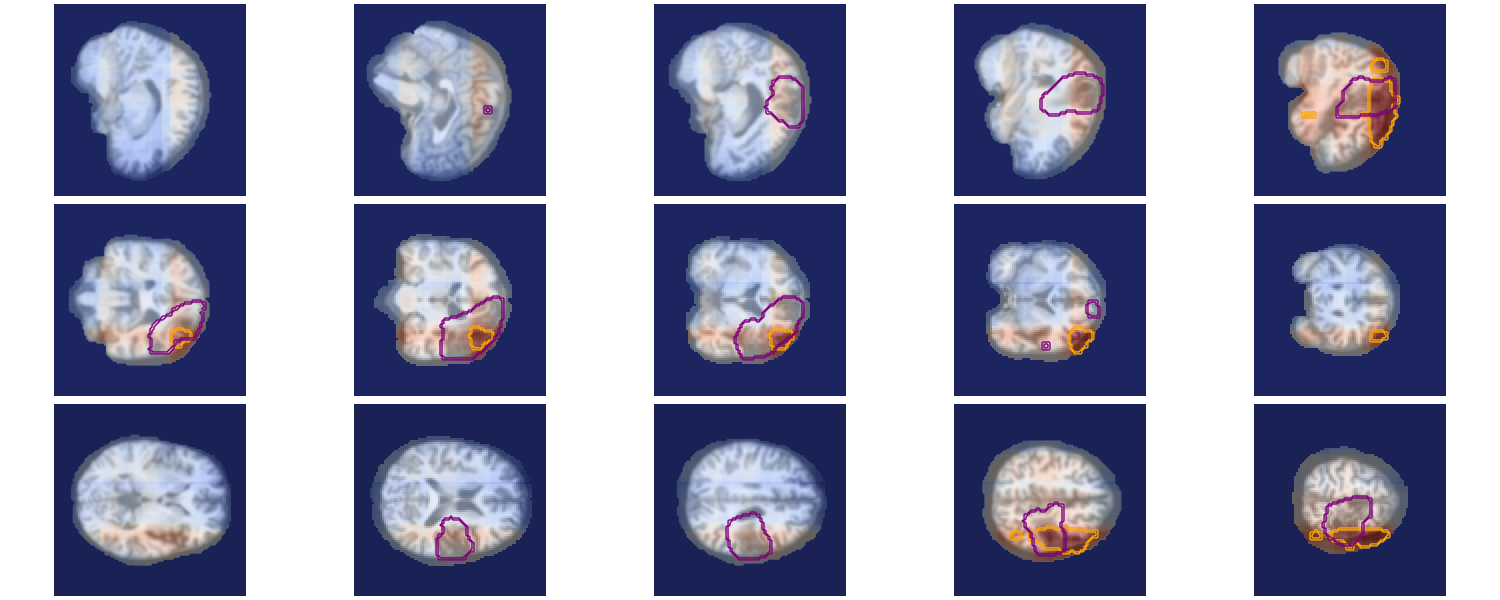}
    \includegraphics[width=0.8\textwidth]{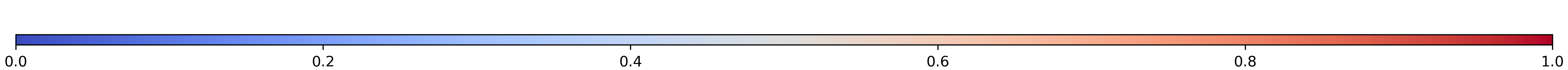}
    \caption{Visualization of the model's explanation for patient A, predicted as stroke with a confidence of $0.9756$. The \textbf{purple} outline represents the annotated lesion location from the dataset, while the \textbf{orange} outline represents the explanation generated by \threedrex. The patient's brain is also overlaid with the responsibility map generated by \threedrex, and the color bar depicts the normalized intensity of feature contributions.}
    \label{fig:overlay}    
\end{figure*}

\begin{figure}[t]
    \centering
    \includegraphics[scale=0.1]{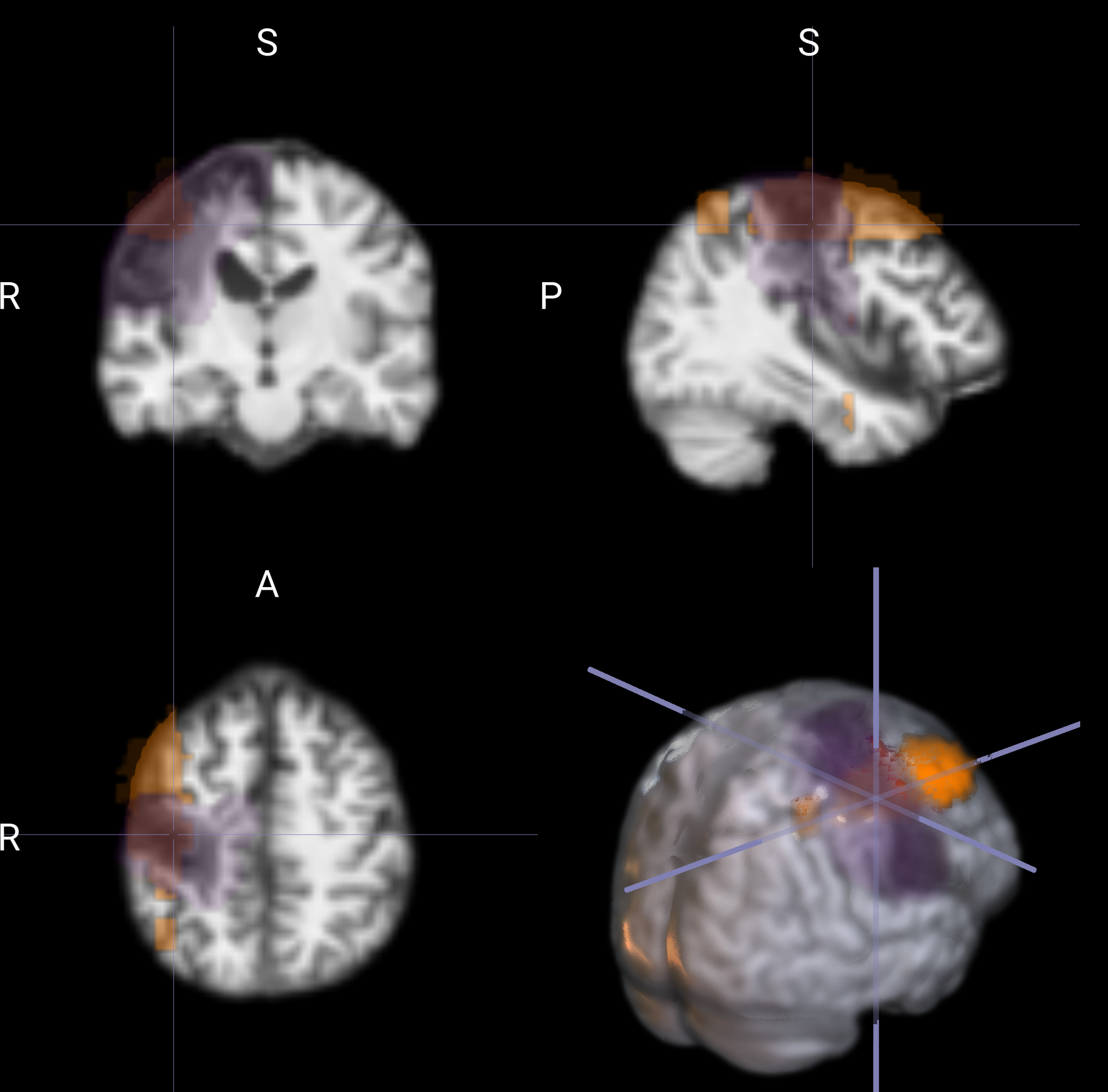}
    \caption{3D rendering of patient A's results showing the patient's brain with overlaid annotated lesion location from the dataset (\textbf{purple}) and 3D-ReX's explanation (\textbf{orange}). This visualization provides a spatial perspective of the explanation.}
    \label{fig:3D}
\end{figure}

\paragraph*{Examples of Results}
\Cref{fig:overlay,fig:3D} illustrate the responsibility map generated by \threedrex superimposed on the MRI scans for patient A. They also outline the explanation extracted from the responsibility map and the lesion location annotation for comparison.

The two visualizations provide complementary perspectives on the model's interpretability.~\Cref{fig:overlay} shows a 2D overlay of the responsibility map and explanation, highlighting the areas of high responsibility in red and areas of low responsibility in blue. The visualization is presented across the axial, sagittal and coronal planes, enabling a comprehensive view of the model's focus within the brain. 

In contrast, \Cref{fig:3D} showcases a 3D rendering of the explanation, offering a volumetric view of the model's focus within the entire structure. This captures the spatial distribution of the explanation compared to the lesion location, allowing for the assessment of deeper or overlapping regions that may not be apparent in the 2D slices. In this case, the explanation, shown in~\Cref{fig:3D}, is positioned perpendicularly in comparison to the lesion in 3D space and focuses on the right hemisphere cortical lesion.
Overall, these visualizations should indicate to clinicians the necessary insight into the model, with different dimensional explanations that can aid in understanding the complex brain structures for stroke diagnosis.

\section{Discussion}\label{sec:discuss}

\paragraph{\textbf{Limitations}}
One significant limitation of the proposed approach is the computational complexity arising from the large search space inherent in constructing responsibility maps for 3D data. Although we mitigate this by only splitting on two axes rather than all three, which reduces the computational burden, further optimization is necessary in time-constrained tasks. Techniques such as parallel processing or batching mutants, where supported by the model, and better sampling strategies could enhance the scalability and efficiency.

We do not claim a comprehensive evaluation of \threedrex.
Numerical evaluation of a large-scale dataset remains to be done. While individual examples and visual comparisons provide insight into our technique, the absence of a systematic, standardized assessment framework over a broader dataset limits generalizability. This issue is further emphasized in a broader call for standardized criteria in clinical explainability evaluation, outlined in \citet{evlauation}. This paper highlights the need for explainability tools to be validated across diverse and clinically relevant datasets and user studies to ensure their utility and reliability in real-world applications. As a next step, we plan to conduct a large-scale evaluation to address this limitation and enhance the robustness of our approach.

\paragraph{\textbf{Future Work}}
We intend to extend \threedrex's capabilities across different medical imaging modalities. We will validate this process with different applications that involve more heterogeneous data and complex pathological features. Another avenue for future work is to further investigate the plausibility of finding a range of different explanations using different occlusion values, such as leveraging a healthy brain MRI to create ``real brain'' occlusions, in a diverse set of contexts and larger models, potentially uncovering novel insight into the model's behavior. 

In cases where the model identifies unexpected features or patterns in the data, we will investigate the use of explanations to disentangle whether the model is learning incorrect features or the model has learned a feature that has biological relevance. This aligns with the broader goal of enabling \xai tools to find explanations that are clinically meaningful but also capable of revealing new scientific knowledge.

Future work concerning \rex's integration with other modes of data input type should include multi-modal models, combining both 3D structural imaging such as CT, MRI \emph{etc.} with genomic, patient history and other details. Multi-modal integration will provide a comprehensive understanding and accurate classification, and being able to produce coherent explanations will help mitigate potential bias and validate model decisions in multimodal contexts.

\section*{Acknowledgments}
Blake, Kelly, and Chockler  were supported in part by the UKRI Trustworthy Autonomous Systems Hub (EP/V00784X/1), the UKRI Strategic Priorities Fund to the UKRI Research Node on Trustworthy Autonomous Systems Governance and Regulation (EP/V026607/1), and CHAI -- EPSRC AI Hub for Causality in Healthcare AI with Real Data (EP/Y028856/1).

\bibliography{main}

\end{document}